\documentclass[twocolumn,showpacs,preprintnumbers,amsmath,amssymb,floatfix]{revtex4}

\usepackage{graphicx}
\usepackage{dcolumn}
\usepackage{bm}
\usepackage{bigints}
\usepackage{xcolor}  

\begin{document}
\title{Hybrid Quantum Systems with Collectively Coupled Spin States:\\Suppression of Decoherence through Spectral Hole Burning}

\author{Dmitry O. Krimer}
\email[]{dmitry.krimer@gmail.com}
\author{Benedikt Hartl}
\author{Stefan Rotter}
\affiliation{Institute for Theoretical Physics, Vienna University of Technology (TU Wien), Wiedner Hauptstra\ss e 8-10/136, A--1040 Vienna, Austria, EU}

\begin{abstract}
Spin ensemble based hybrid quantum systems suffer from a significant degree of decoherence resulting from the inhomogeneous broadening of the spin transition frequencies in the ensemble. We demonstrate that this strongly restrictive drawback can  be overcome simply by burning two narrow spectral holes in the spin spectral density at judiciously chosen frequencies. Using this procedure we find an increase of the coherence time by more than an order of magnitude as compared to the case without hole burning. Our findings pave the way for the practical use of these hybrid quantum systems for the processing of quantum information.
\end{abstract}

\maketitle

\begin{figure}[b!]
\includegraphics[angle=0,width=.85\columnwidth]{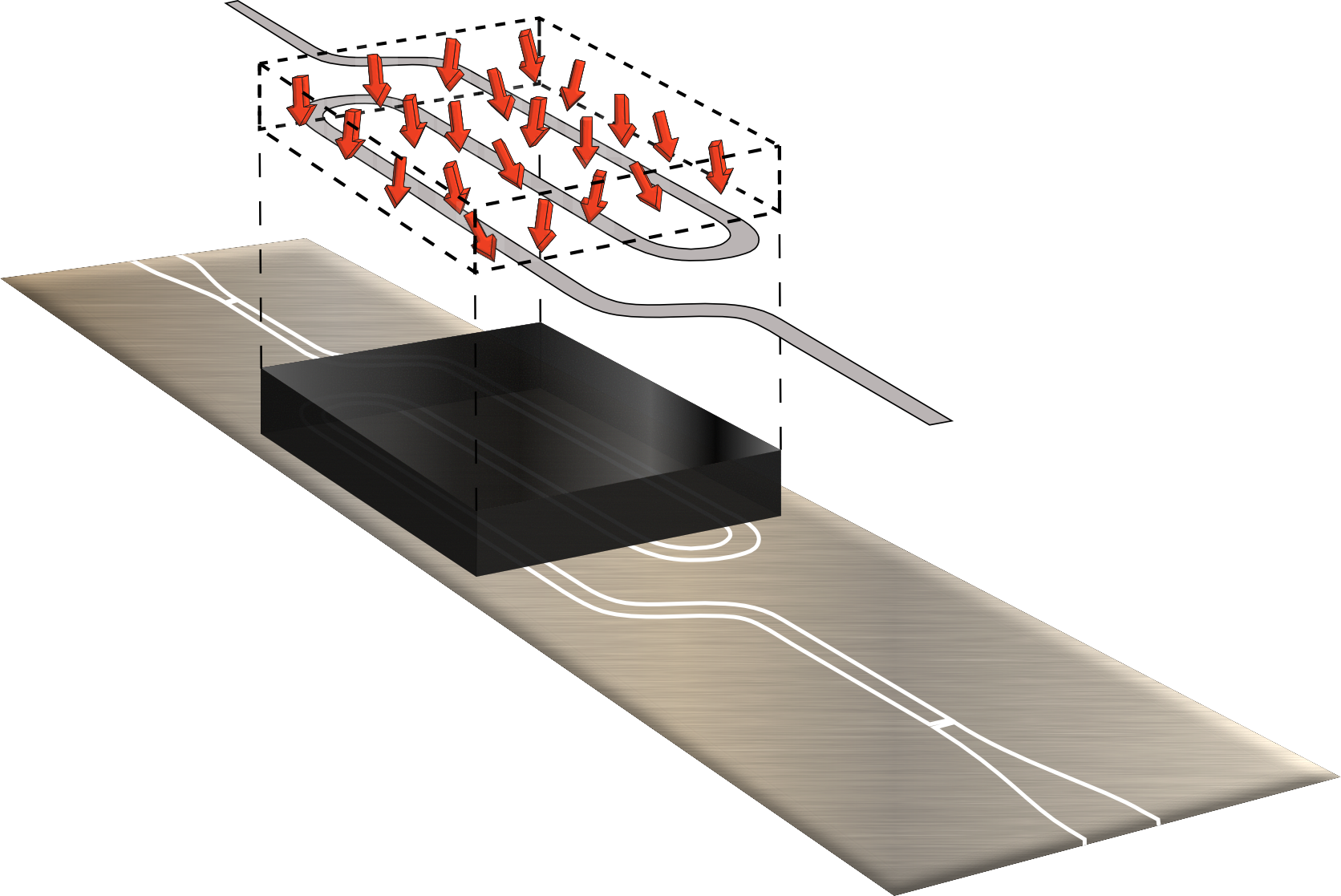}
\caption{(Color online) Sketch of the studied hybrid quantum system: a synthetic diamond (black) containing a spin ensemble (red arrows) coupled to a transmission-line resonator (curved gray line) confining the electromagnetic field to a small volume.}
\label{Fig1_ensemble_cavity}
\end{figure}

Hybrid quantum circuits that conflate the advantages of different physical systems to achieve new device functionalities have recently shifted to the center of attention \cite{Xiang2013}. This is largely because a new generation of experiments \cite{Amsuess2011, Sandner2012, Kubo2010, Kubo2011, Kubo2012, Schuster2010, Bushev2010, Zhu2011, Probst2013, Nature2014} lends encouraging plausibility to the vision of using such hybrid device concepts to reliably store and manipulate quantum information \cite{Rabl2006, Tordrup2008, Verdu2009, Petrosyan2009, Imamoglu1999, Wesenberg2009}. In particular, the recent achievements in strongly coupling large spin ensembles to superconducting microwave cavities \cite{Amsuess2011, Kubo2010, Kubo2011, Sandner2012, Kubo2012, Nature2014} hold promise for combining many of the advantageous features of microwave technology with the long spin coherence times found, e.g., in crystallographic defects of diamond.
  
Whereas the collective coupling to a whole ensemble of spins is the key to reach the strong-coupling
limit, the ensemble generally comes with the downside of being inhomogeneously broadened, i.e., the transition frequencies between different spin levels are slightly different for each spin. As it turns out, the decoherence resultant from this broadening is currently the major bottleneck for the processing of quantum information in these hybrid quantum systems. First attempts at resolving this problem have meanwhile been put forward: On the one hand, it was shown that the decoherence is naturally suppressed for very strong coupling when the spectral spin distribution realized by the ensemble falls off sufficiently fast in its tails. Signatures of this so-called ``cavity protection effect'' \cite{Kurucz2011,Diniz2011}, have meanwhile also been observed experimentally \cite{Nature2014,KPMS14}. To fully bring to bear the potential of this effect requires, however, to go to very high values of the coupling strength, which are presently difficult to reach experimentally. On the other hand, sophisticated  concepts for the spectral engineering of the spin density profile have been proposed \cite{Bensky2012,Jelezko2012}. These concepts rely, however, on a strong modification of the intrinsically predefined density profile that is again very challenging to implement experimentally. In this Letter, we present a method that circumvents the problems of both approaches by building on a very elementary concept that requires only a considerably reduced experimental effort. Specifically, we demonstrate that the burning of two judiciously placed spectral holes in the spin distribution suffices to drastically increase the coherence properties of the hybrid spin-cavity system. From the viewpoint of quantum control our approach constitutes a new and efficient strategy to stabilize Rabi oscillations in the strong-coupling limit of cavity QED \cite{Vijay2012,Carmele2013,Kabuss2015}. Suppressing the detrimental influence of inhomogeneous broadening, as suggested in our work, could also prove to be a key element for the realization of ultra-narrow linewidth lasers \cite{Meiser2009,Bohnet2012}.

To connect our theoretical work directly with the experiment we will study in the following the recently implemented case of a superconducting microwave resonator strongly coupled to an ensemble of negatively charged nitrogen-vacancy centers in a diamond (see Fig.~\ref{Fig1_ensemble_cavity}) \cite{Amsuess2011, Sandner2012, Nature2014, KPMS14}. Our starting point is the Tavis-Cummings Hamiltonian ($\hbar=1$) \cite{Tavis68}, which describes the dynamics of a single-mode cavity coupled to a spin ensemble in the dipole and rotating-wave approximation,
\begin{eqnarray}
{\cal H}&=&\omega_ca^{\dagger}a+\frac{1}{2}\sum_j^N\omega_j\sigma_j^z+\text{i}\sum_j^N\left[g_j\sigma_j^-a^{\dagger}-g_j^*\sigma_j^+a\right]-
\nonumber\\
&&\text{i}\left[\eta(t) a^{\dagger}\text{e}^{-\text{i}\omega t}-\eta(t)^* a\text{e}^{\text{i}\omega t}\right]\,.
\label{Hamilt_fun}
\end{eqnarray}
Here $\sigma_j^+,\,\sigma_j^-,\,\sigma_j^z$ are the Pauli operators associated with the individual spins of frequency $\omega_j$. Each spin is coupled with a strength $g_j$ to the single cavity mode of frequency $\omega_c$, in which photons are created and annihilated through the operators $a^{\dag}$ and $a$. The probing electromagnetic field injected into the cavity is characterized by its carrier frequency $\omega$ and by the amplitude $\eta(t)$. 

Next, we derive the semiclassical equations of motion using the Holstein-Primakoff-approximation \cite{Primakoff1939} (implying that the condition $\langle \sigma_k^z \rangle \approx -1$ always holds), the rotating-wave approximation and neglecting the dipole-dipole interaction between spins. With these simplifications, which are well justified for the experiments \cite{Nature2014, KPMS14} operating at low input powers of an incoming signal, the equations for $A(t)\equiv \langle a(t)\rangle$ and $B_j(t)\equiv\langle\sigma_j^-(t)\rangle$ acquire the following form (in the $\omega$-rotating frame),
\begin{subequations}
\begin{eqnarray}
\label{Eq_a_Volt}
&&\!\!\!\!\!\!\!\!\!\!\!\!\!\!\!\! \dot{A}(t)= -\left[\kappa+i(\omega_c-\omega)\right]A(t) + \sum_j
g_j  B_j(t)-\eta(t), \\
\label{Eq_bk_Volt}
&&\!\!\!\!\!\!\!\!\!\!\!\!\!\!\!\!\dot{B}_j(t) = -\left[\gamma+i(\omega_j-\omega) \right]B_j(t) - g_j A(t),
\end{eqnarray}
\end{subequations}
where $\kappa$, $\gamma$ are the dissipative cavity and spin losses, respectively. 

Large spin ensembles ($N \sim10^{12}$ in \cite{Nature2014, KPMS14}) are best described by the continuum limit of the normalized spectral density $\rho(\omega)=\sum_j^N g_j^2 \delta(\omega-\omega_j)/\Omega^2$. Here $\Omega=(\sum_j^Ng_j^2)^{1/2}$ is an effective coupling strength which is enhanced by a factor of $\sqrt{N}$ as compared to a single coupling strength, $g_j$, so that $\Omega$ can reach the values necessary for the realization of the strong coupling regime. The inhomogeneous broadening of the spin frequencies $\omega_j$ and coupling strengths $g_j$ then lead to a finite-width distribution $\rho(\omega)$ centered around a certain mean frequency $\omega_s$. The specific shape of this spectral density $\rho(\omega)$ can typically be determined by a careful comparison with the experiment based on stationary \cite{Sandner2012}  or dynamical \cite{Nature2014} transmission measurements. In the following we will use the same parameters as in \cite{Nature2014, KPMS14} taking a $q$-Gaussian distribution \cite{Sandner2012} for $\rho(\omega)$ centered around $\omega_s/2\pi=2.6915$\,GHz, a full-width at half-maximum of $\gamma_q/2\pi=9.44$\,MHz and a $q$-parameter of $1.39$. The cavity decay rate, $\kappa/2\pi=0.4$\,MHz (half-width at half-maximum) and the coupling strength $\Omega/2\pi=8.56$ MHz. 

The starting point for our strategy is the insight that the non-Markovian dynamics of the spin system, which is described by $\rho(\omega)$ and strongly coupled to the cavity mode, can be accurately modeled by an integral Volterra equation for the cavity amplitude $A(t)$ (see Eq.~(\ref{Eq_rigor}) below and \cite{Nature2014,KPMS14}). The latter includes a memory-kernel, which is responsible for the non-Markovian feedback of the spin ensemble on the cavity, so that the cavity amplitude at time $t$ depends on all previous events $\tau<t$. By performing a Laplace transform of this Volterra equation \cite{KPMS14} or by carrying out a stationary transmission analysis \cite{Diniz2011,Kurucz2011}, the total rate of decoherence turns out to be $\Gamma\approx\kappa+\pi\Omega^2\rho(\omega_s\pm\Omega)$ in the limit of large coupling strengths, $\Omega > \Gamma$ and $\gamma\to 0$. The value of $\Gamma$ is thus determined by the spin density $\rho(\omega)$, evaluated close to the maxima of the two polaritonic peaks, $\omega=\omega_s\pm\Omega$, split by the Rabi frequency $\Omega_R\approx 2\Omega$ due to strong coupling. Our approach is now to take this relation literally, which is tantamount to saying that the decoherence rate $\Gamma$ can be strongly suppressed by burning two spectral holes into the spin distribution $\rho(\omega)$ {\it right at these two positions}, $\omega_h=\omega_s\pm\Omega$, such that $\rho( \omega_h)=0$. The width of the holes $\Delta_h$ should be very small, such as to remove only a negligible fraction of the spins by the hole burning. On the other hand, $\Delta_h$ is limited from below by the spin dissipation rate,  $\Delta_h>\gamma$.

\begin{figure}
\includegraphics[angle=0,width=1.\columnwidth]{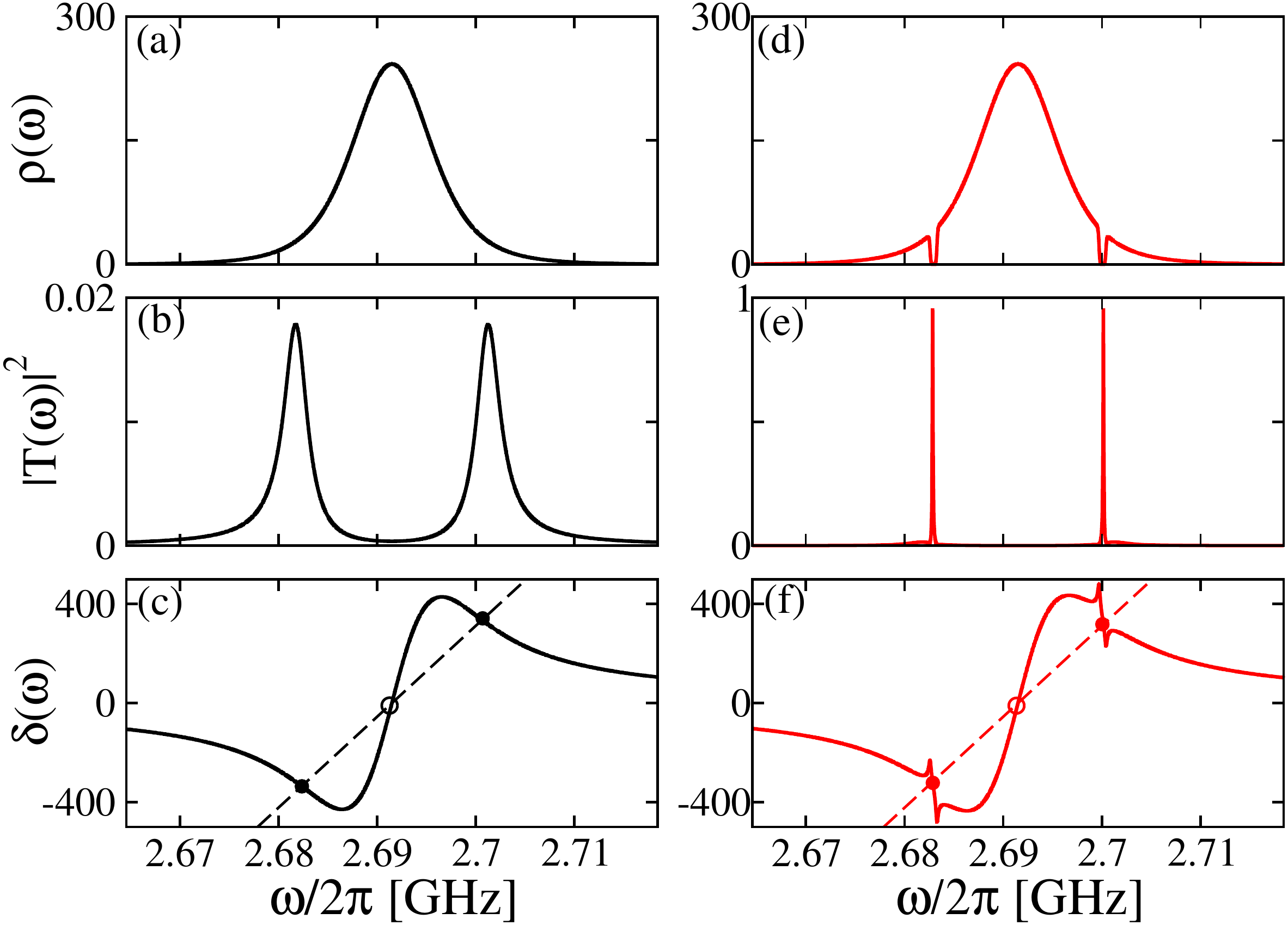}
\vspace*{-0.6cm}
\caption{(Color online) Comparison of the cavity coupled to the inhomogeneously broadened spin ensemble without and with hole burning in the spin density profile (left and right panels, respectively). Top row: The $q$-Gaussian spin density distribution, $\rho(\omega)$, without and with hole burning at $\omega_h=\omega_s\pm\Omega$. Both holes are of equal width, $\Delta_h/2\pi=0.7$\,MHz, and have a Fermi-Dirac profile. Middle row: Transmission $T(\omega)$ without and with hole burning in $\rho(\omega)$ (note different $y$-axes scale). Bottom row: The corresponding nonlinear Lamb shift $\delta(\omega)$. Filled circles label resonance values $\omega_r$ of the transmission $T(\omega)$ occurring at the intersections between the Lamb shift $\delta(\omega)$ and the dashed line $(\omega-\omega_c)/\Omega^2$. At empty circles such intersections are non-resonant (see text).}
\label{fig_Transmission_Lamb_shift_FD_0p7MHz}
\end{figure}

To demonstrate the efficiency of our approach explicitly, we first perform a stationary analysis [$\dot{A}(t)=\dot{B}_k(t)=0$] of the transmission $T(\omega)$ through the microwave resonator as a function of the probing frequency $\omega$. This quantity, which is directly accessible in the experiment \cite{Nature2014,KPMS14}, provides direct access to the occupation amplitude of the cavity [$T(\omega)\propto A(\omega)$]. Assuming $\gamma \rightarrow 0$, the transmission $T(\omega)$ acquires the following form,
\begin{eqnarray}
\label{Eq_Phis_34}
T(\omega)=\dfrac{i\kappa}
{\omega\!-\!\omega_c\!-\!\Omega^2 \delta(\omega)+i[\kappa+\pi\Omega^2\rho(\omega)]}\,.
\end{eqnarray}
This expression is normalized such as to reach the maximum possible value max$(|T(\omega)|)=1$ for suitably chosen $\omega$, $\kappa$,  and $\rho(\omega)$. The real function $\delta(\omega)$ is the nonlinear Lamb shift \cite{KLRT14} defined as
\begin{eqnarray}
\delta(\omega)=\mathcal{P}\int_0^{\infty}\dfrac{d\tilde\omega \rho(\tilde\omega)}{\omega\!-\!
\tilde\omega}\!,
\label{Eq_Lamb_shift}
\end{eqnarray}
where $\mathcal{P}$ stands for the Cauchy principal value. In the reference case taken from the experiment \cite{Nature2014,KPMS14}, $\rho(\omega)$ has no holes, see Fig.~\ref{fig_Transmission_Lamb_shift_FD_0p7MHz}(a), and the transmission $|T(\omega)|^2$ displays the well-resolved double-peak structure typical for the strong-coupling regime, see Fig.~\ref{fig_Transmission_Lamb_shift_FD_0p7MHz}(b). If we now burn two narrow holes into the spin density at the relevant positions $\omega_h= \omega_s\pm\Omega$, see Fig.~\ref{fig_Transmission_Lamb_shift_FD_0p7MHz}(d), and reevaluate $|T(\omega)|^2$ we observe a more than fiftyfold increase in the corresponding transmission peak values, see 
Fig.~\ref{fig_Transmission_Lamb_shift_FD_0p7MHz}(e). This dramatic change is all the more
surprising considering that the relative number of spins removed from $\rho(\omega)$
through the hole burning is less than $3\%$.

To understand this behavior it is best to analyze the real and imaginary parts of the denominator of $T(\omega)$, see Eq.~(\ref{Eq_Phis_34}). 
For the observed transmission resonances at $\omega=\omega_r$ with a maximum value of $T(\omega_r)=1$ to occur, two conditions are satisfied simultaneously: 
(i) $(\omega_r-\omega_c)/{\Omega^2}= \delta( \omega_r)$ and (ii) $\rho(\omega_r)=0$. Consider first condition (i): In the reference case without holes, see Fig.~\ref{fig_Transmission_Lamb_shift_FD_0p7MHz}(c), the nonlinear Lamb shift $\delta(\omega)$ 
displays rather smooth variations in the vicinity of the resonant frequencies $\omega_r$, determined by the intersection of $\delta(\omega)$ and a straight line $(\omega-\omega_c)/\Omega^2$. In contrast, for the case with hole burning, see Fig.~\ref{fig_Transmission_Lamb_shift_FD_0p7MHz}(f), $\delta(\omega)$ exhibits rapid variations around the two resonance points within a very narrow spectral interval. As a consequence, the resultant  transmission peaks become substantially sharper. Due to the second condition (ii) they also dramatically increase in height. Note, that no resonance occurs at $\omega=\omega_c$ because $\rho(\omega)$ has a maximum at this point and condition (ii) is strongly violated, see Fig.~\ref{fig_Transmission_Lamb_shift_FD_0p7MHz}(c),(f). A close examination of the structure of $T(\omega)$ shows, furthermore, that the narrow transmission peaks resultant from the hole burning do not replace the broad polaritonic peaks present in the reference case, but rather get to sit on top of them, see Fig.~\ref{fig_TRM_MAP_W14_FD}(a). As will be seen below, the different resonance widths in $T(\omega)$ set  two different time scales in the dynamics with, in particular, the sharp peaks in the transmission giving rise to an asymptotically slowly decaying dynamics with a strongly suppressed decoherence.
\begin{figure}
\includegraphics[angle=0,width=0.9\columnwidth]{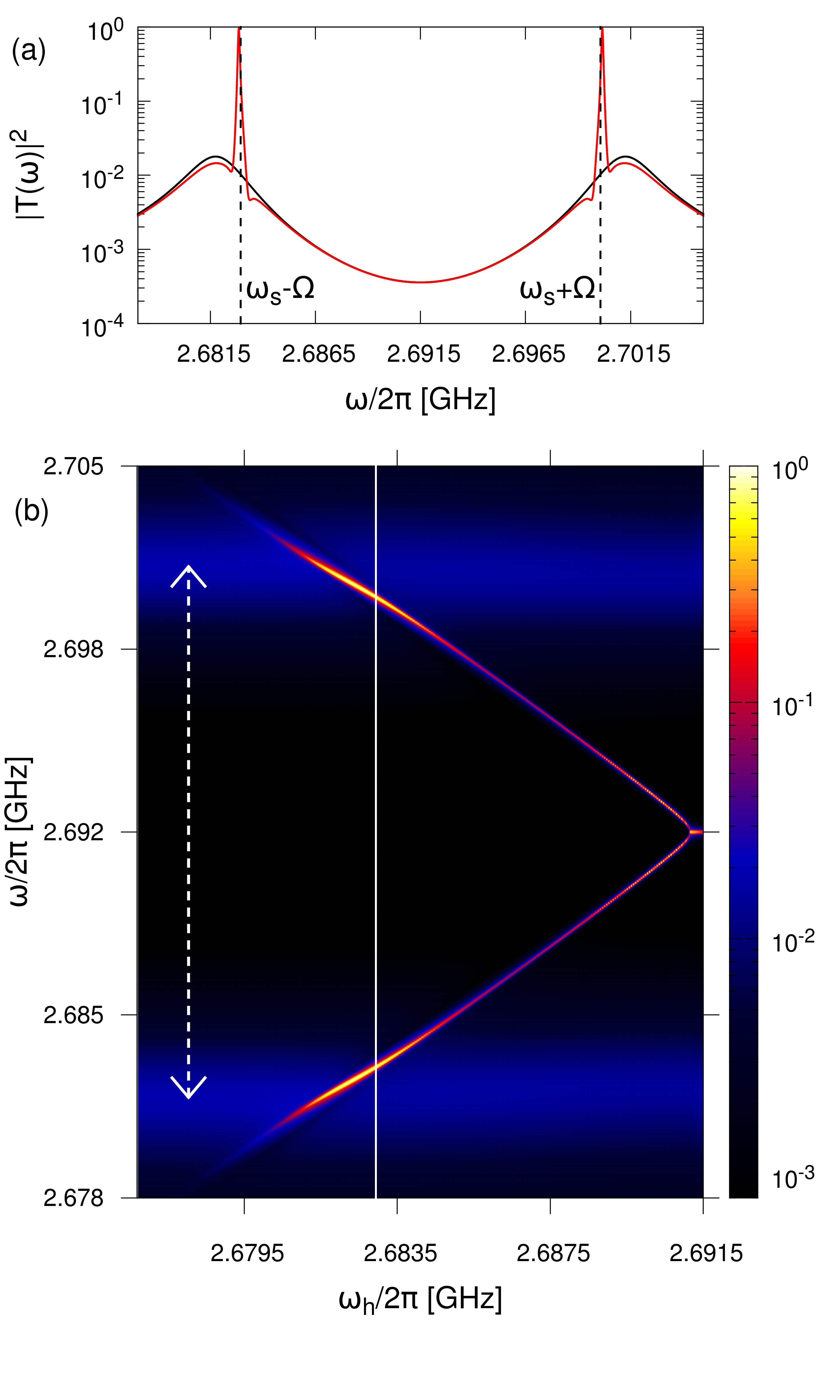}
\vspace*{-1.cm}
\caption{(Color online) Transmission through the cavity $T(\omega)$ versus probe frequency $\omega$ for different locations of the holes, $\omega_h$, in the spin density profile, $\rho(\omega)$ (the width of the holes is $\Delta_h/2\pi=0.7$\,MHz). (a) Red (gray) curve: $|T(\omega)|^2$ in lin-log scale versus $\omega$ for $\omega_h=\omega_s\pm\Omega$. Black curve: Transmission in the absence of hole burning. (b) Yellow (light gray) areas mark the most pronounced peaks in $|T(\omega)|^2$ in the presence of hole burning. Blue (gray) areas stand for the secondary polaritonic peaks which stem from the case without hole burning. Dashed arrows designate the distance 
$\Omega_R$ between polaritonic peaks. The white vertical cut corresponds to the transmission shown in (a).}
\label{fig_TRM_MAP_W14_FD}
\end{figure}
\begin{figure}
\includegraphics[angle=0,width=1.\columnwidth]{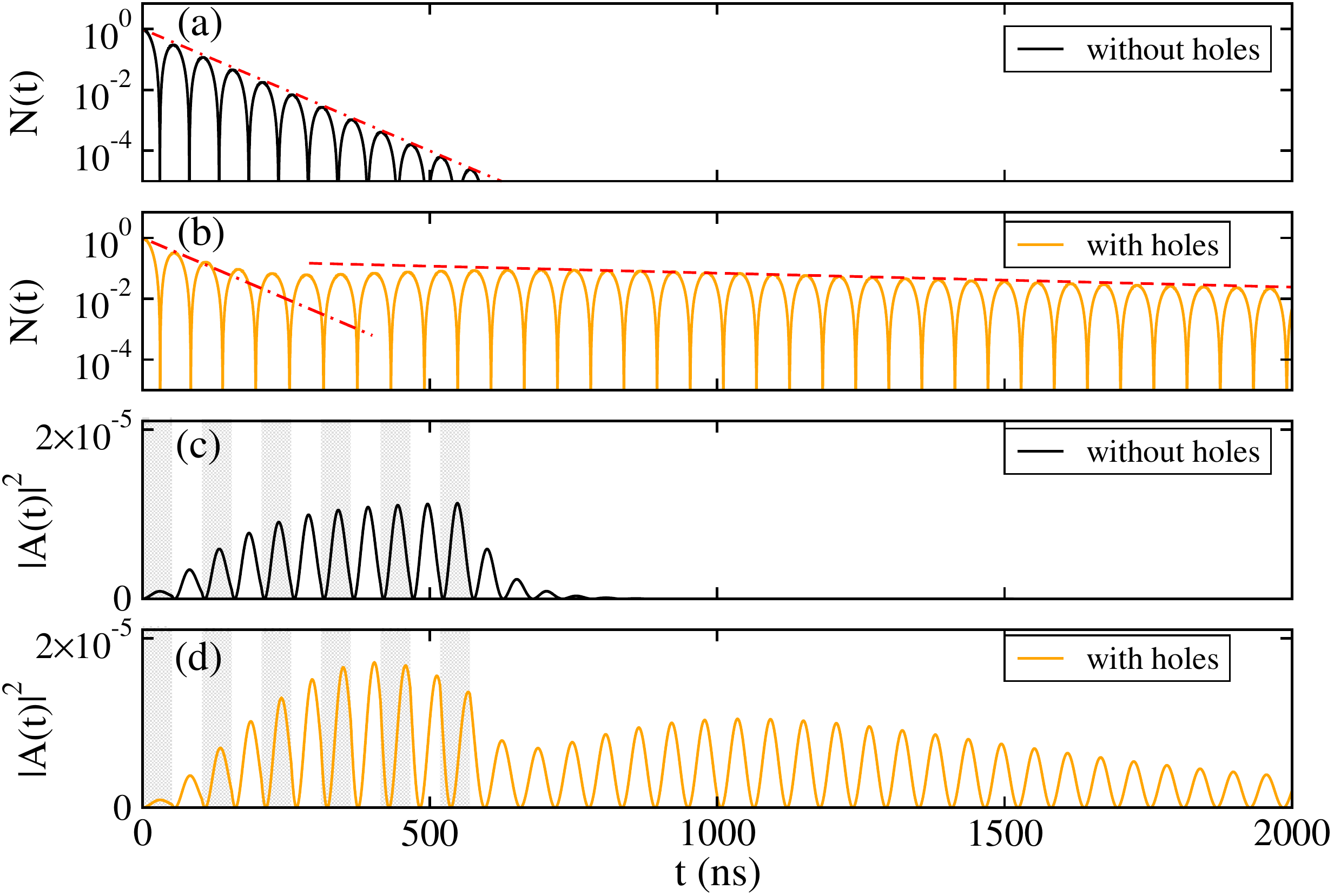}
\vspace*{-0.5cm}
\caption{(Color online) (a),(b): Decay of the cavity occupation $N(t)=\langle 1,\downarrow\!\!|a^{\dagger}(t) a(t)|1,\downarrow \rangle$ from the initial state, for which a single photon with frequency $\omega_c$ resides in the cavity and all spins are unexcited. The asymptotic decay $C e^{-\Gamma t}$ with and without hole burning (see red lines) is determined by the constants  $\Gamma/2\pi=3$\,MHz in (a) and a drastically reduced $\Gamma=0.42 \kappa=2\pi \cdot 0.17$\,MHz in (b). (c),(d): Dynamics of $|A(t)|^2$ under the action of eleven successive rectangular microwave pulses of duration corresponding to the Rabi period, $\tau=2\pi/\Omega_R=52$\,ns, phase-switched by $\pi$ (every second pulse is shown as a vertical gray bar).  Also here the asymptotic decay is much slower due to the presence of the holes. In all panels the holes in $\rho(\omega)$ have a width $\Delta_h/2\pi=1.4$\,MHz and are burnt at $t=0$ at $\omega_h=\omega_s\pm\Omega$. }
\label{fig_dynamics}
\end{figure}

To explore whether the narrow holes we burnt into the spectral spin distribution at $\omega_h=\omega_s\pm\Omega$ have, indeed, the optimal location, we now also test all possible other hole positions symmetrically placed around the maximum of $\rho(\omega)$ at $\omega=\omega_s$. In Fig.~\ref{fig_TRM_MAP_W14_FD}(b) we present the numerical results for $T(\omega)$ as a function of the probe frequency $\omega$ and of different hole locations $\omega_h=\omega_s\pm\bar{\omega}$ scanned between $\bar{\omega}=0$ and $\bar{\omega}=16$\,MHz: While for large hole spacings  ($\bar{\omega}\gtrsim 11.5$\,MHz) the effect of holes is negligible, in the interval $0.8$\,MHz\,$\lesssim\! \bar{\omega}\!\lesssim 11.5$\,MHz we always find two sharp peaks superimposed on the two polaritonic peaks approximately at the hole positions. Close to $\omega_h=\omega_s\pm\Omega$ these peaks are most pronounced and reach unity. In the limit when the holes are burnt very close together ($\bar{\omega}\lesssim 0.8$\,MHz) the sharp peaks merge into a single one, located directly at the central frequency $\omega_s$ with a transmission maximum reaching again unity in the limit of $\omega_h\rightarrow \omega_s$ [see the yellow cusp in Fig.~\ref{fig_TRM_MAP_W14_FD}(b)]. Using the symmetry of $\rho(\omega)$ with respect to $\omega_s$, this behavior can also be proven analytically (not shown). To check the robustness of our method we also tested different functional forms for the hole profiles (Fermi-Dirac, $q$-Gaussian and rectangular distributions) and found qualitatively similar results to the Fermi-Dirac form employed for all of the above figures.   

To reach our ultimate goal of understanding the influence of the spectral hole burning on the resultant dynamics, we now study the time evolution of $A(t)$ explicitly for the resonant case $\omega=\omega_c=\omega_s$. The expression for the corresponding Volterra equation can be derived from Eqs.~(\ref{Eq_a_Volt}, \ref{Eq_bk_Volt}) (see \cite{KPMS14} for details),
\begin{eqnarray}
\label{Eq_rigor}
&&\dot A(t)=-\kappa A(t)-
\\
&&\Omega^2 \int d\omega \rho(\omega) \int\limits_{0}^t d\tau
e^{-i(\omega-\omega_c-i\gamma)(t-\tau)}A(\tau)-\eta(t)\,.\nonumber
\end{eqnarray}
To prove that our predictions are valid not only in the semiclassical but also in the quantum case, we consider the case when all spins are initially in the ground state and the cavity mode $a$ contains initially a single photon, $|1,\downarrow \rangle$. It can be shown that the probability for a photon to reside in the cavity at time $t>0$, $N(t)=\langle 1,\downarrow\!\!|a^{\dagger}(t) a(t)|1,\downarrow \rangle$, reduces to $N(t)=|\langle 0, \downarrow\!\!| a(t)|1,\downarrow \rangle|^2=|A(t)|^2$, where $A(t)$ is the solution of Eq.~(\ref{Eq_rigor}) with the initial condition $A(t=0)=1$ (external drive $\eta(t)=0$). For the case without hole burning this solution is represented by the damped Rabi oscillations [see Fig.~\ref{fig_dynamics}(a)] found already previously \cite{Nature2014,KPMS14}. By burning narrow holes in $\rho(\omega)$ at $\omega_h=\omega_s\pm\Omega$ (immediately before $t=0$), we observe very similar transient dynamics, which is followed, however, by a crossover to Rabi oscillations with a much slower asymptotic decay [see Fig.~\ref{fig_dynamics}(b)]. Quite remarkably, the total decay rate $\Gamma$ in this asymptotic time limit can even be substantially smaller than the cavity decay rate $\kappa$ alone. This is all the more surprizing since $\kappa$ was identified as the minimally reachable value for $\Gamma$ in recent studies on the cavity protection effect \cite{Diniz2011,Nature2014,KPMS14}. Apparently a new type of physics is at work here: Although the system is in the strong coupling regime, the two spectral holes slow down the leakage of the energy stored in the spin ensemble back into the cavity. In particular, when being even slower than the inverse of the cavity decay rate $\kappa$, this sets a new global time scale for  $\Gamma$, corresponding to the width of the sharp resonance peaks which we identified before in Fig.~\ref{fig_TRM_MAP_W14_FD}(a). From the mathematical point of view such a slow asymptotic behavior is associated with the contribution of two poles in the Laplace transform of Eq.~(\ref{Eq_rigor}) \cite{KPMS14}, which appear when the holes in $\rho(\omega)$ reach a critical depth. The pole contributions also stabilize the long-time behavior when the holes are shifted away from the polaritonic peaks [see Fig.~\ref{fig_TRM_MAP_W14_FD}(b)], but the optimal hole positions remain close to the polaritonic peaks. Note that despite the considerable photon loss $(N(t) \ll 1)$ for long times the phase coherence is very well preserved here, a clear signature of which is the stable  form of the Rabi oscillations. In this way a high ``visibility'' can be achieved, as required for the efficient processing of quantum information \cite{Gisin2008}.

To demonstrate the efficiency of the hole burning effect also for quantum control schemes, we pump the cavity by a sequence of $\pi$ phase-switched rectangular pulses, each with a duration corresponding to the Rabi period, $\tau=2\pi/\Omega_R$ and a carrier frequency $\omega=\omega_c=\omega_s$. As shown in \cite{Nature2014}, this procedure is very well suited to feed energy into the strongly coupled cavity-spin system, leading to giant oscillations of both spin and cavity amplitude [see left parts of Fig.~\ref{fig_dynamics}(c,d)]. Not only do we observe that these driven oscillations are more pronounced when burning holes at $\omega_h=\omega_s\pm\Omega$, but we find, in particular, that the Rabi relaxation oscillations setting in after switching off the driving field are dramatically more long-lived than in the case without holes [compare right parts of Fig.~\ref{fig_dynamics}(c,d)]. These results confirm the robustness as well as the general applicability of our approach for various coherent-control schemes in the strong-coupling regime of cavity QED.

In summary, we present an efficient method to suppress the decoherence in a single-mode cavity strongly coupled to an inhomogeneously broadened spin ensemble. By burning narrow spectral holes in the spin density at judiciously chosen positions the total decay rate is dramatically decreased to values that may even lie below the dissipation rate of the bare cavity. Experimentally, our approach can be implemented by exposing the cavity to high-intensity microwave signals with spectral components near the desired hole positions.  Due to the strong driving the spins at these frequencies will equally populate their ground and excited state and will thus be effectively removed from the coupling process with the cavity.

\section{Acknowledgements} We thank S. Putz, R. Ams{\"u}ss and J. Majer for suggesting us to work on this topic and for pointing out other spectral hole burning technique as studied in [21]. Support by the Austrian Science Fund (FWF) through Project No. F49-P10 (SFB NextLite) is gratefully acknowledged.

\end{document}